# Electronic nematic normal and superconducting state in electron-doped copper-oxide superconductors

J.Y. Shen[1, 2, 3, 4], G.F. Chen[1, 2, 3, 4], Y.C. Zhang[1, 2, 3, 4], G.Y. Xi[1, 2, 3, 4], J.Y. He[1, 2, 3], X.B. Cheng[1, 2, 3, 4], J. Wu[1, 2, 3]*

[1]*Department of Physics, School of Science, Westlake University, Hangzhou 310024, China*

[2]*Research Center for Industries of the Future, Westlake University, Hangzhou31 0024, China*

[3]*Key Laboratory for Quantum Materials of Zhejiang Province, School of Science, Westlake University, Hangzhou, 310024, China*

[4]*School of Physics, Zhejiang University, Hangzhou 310027, China*

*\*Author to whom correspondence should be addressed: wujie@westlake.edu.cn*

**Abstract:**

**The similarities and differences between hole- and electron-doped cuprates are central to studies of high-temperature superconductivity. While electronic nematicity is found to be pervasive in hole-doped cuprates, iron-based superconductors, and other unconventional superconductors, evidence for electronic nematicity in electron-doped cuprates remains elusive. Here, we discover that the normal state of electron-doped $Sr_{0.9}La_{0.1}CuO_2$ (SLCO) is nematic by the angle-resolved resistivity (ARR) method and the uncovered ground state at zero temperature is also nematic when superconductivity is suppressed by an applied magnetic field. As we deliberately change the substrate from tetragonal $KTaO_3(001)$ (KTO) to orthorhombic $GdScO_3(110)$ (GSO), the nematic director of SLCO is pinned by the epitaxial strain but the nematic amplitude remains roughly the same, implying that the nematicity originates**

**from electron-electron correlations. The nematicity is significantly enhanced by the presence of superconducting fluctuations and its amplitude increases appreciably as the effective doping level of SLCO is lowered from optimal to underdoped. Thus, electronic nematicity is intrinsic to high-temperature superconductors regardless of differences in the structural and electronic configurations corresponding to hole or electron doping.**

Cuprates can be doped either with holes or electrons to become superconducting[1,2]. The comparison between these two may unravel the essential ingredients necessary for high-temperature superconductivity. Aside from the undoped antiferromagnetic phase and the superconducting dome, the normal state of electron-doped cuprates shows both similarities and asymmetries with that of their hole-doped counterparts. For instance, the strange metal behavior was shown to be correlated with the superfluid stiffness, and the scaling law was discovered to be universal for electron- and hole-doped cuprates[3-5]. On the other side, the pseudogap manifests itself through a variety of probes in hole-doped cuprates, e.g., electric transport[6,7], scanning tunneling microscopy[8], etc., but it is absent in electron-doped cuprates[3]. These findings should be contrasted with the finding[1] that doped electrons reside at copper atoms to form Cu $d^{10}$ configuration while doped holes reside at oxygen atoms and are coupled with Cu $d^9$ electrons to form Zhang-Rice singlet. These results raise the key question on which of the abundant charge and spin orders[9] in the normal state of cuprates are not incidental but generically related to superconductivity.

Electronic nematicity[10-12], which denotes the spontaneous breaking of in-plane rotational symmetry, is found to be pervasive among unconventional superconductors, such as iron-based superconductors[13-15], heavy-Fermion superconductors[16], and twisted-layer graphene[17]. For hole-doped cuprates, the nematicity was discovered by complementary methodologies, e.g., electric transport[18-22], scanning tunneling microscopy[23], angle-resolved photoemission spectroscopy[24,25], optical measurements[26,27], etc. Moreover, compelling evidence from electric transport studies of hole-doped La$_{2-x}$Sr$_x$CuO$_4$ (LSCO)[20,22,28] elucidated that the rotational symmetry and the $C_4$ symmetry of the in-plane lattice are spontaneously reduced to $C_2$ symmetry, and the normal state above the superconducting dome is all electronic nematic, from underdoped to overdoped, from room temperature to superconducting temperature $T_c$. Thus, for hole-doped cuprates, superconductivity emerges out of the electronic nematic state and vice versa.

It is vital to examine whether the same picture applies to electron-doped cuprates. Compared to their hole-doped counterparts, electron-doped cuprates are less well explored, and no evidence of nematicity has been reported until now. In this paper, we systematically study the longitudinal and transverse resistivity along different in-plane directions of $CuO_2$ plane in electron-doped SLCO thin films. We discover unambiguous evidence demonstrating electronic nematicity in the normal state from room temperature down to $T_c$ as well as enhanced nematicity due to superconducting fluctuations around $T_c$.

The synthesis of infinite-layer SLCO film[29-31] (Fig. 1a) is technically challenging and demands delicate control over synthesis conditions. The SLCO films were deposited onto KTO or GSO substrates by pulsed laser deposition and were annealed at 525 °C in vacuum for 25 minutes to remove excess interstitial oxygen. The growth recipe is the result of our continuous endeavors devoted to the growth-characterization cycles to produce single-crystalline SLCO films with the right stoichiometry (Figs. 1b-1f). The crystal structure of the synthesized SLCO films was confirmed by X-ray diffraction (XRD), with both SLCO/KTO and SLCO/GSO displaying distinct peaks corresponding to the SLCO(002) diffraction peak (Figs. 1b and 1c). The fringe peaks in XRD spectra are evidence of atomically sharp SLCO/KTO and SLCO/GSO interfaces. The atomic force microscopy (AFM) shows that the root-mean-square (r.m.s.) surface roughness of SLCO/KTO is only 0.44 nm (Fig. 1d), which is comparable to the height of the atomic step in the KTO substrate, 0.40 nm. To characterize the superconducting state of SLCO, we utilized the two-coil mutual inductance method[5, 32-34]. An *ac* magnetic field with 30 kHz frequency is generated by a drive coil and then penetrates the SLCO film before it reaches the pick-up coil. Due to the Meissner effect, the magnetic flux in the pick-up coil, which is proportional to the induced in-phase voltage Re$V_p$, drops substantially as the SLCO film enters the superconducting state (Fig. 1e). The penetration depth $\lambda(T)$ and the superfluid stiffness $\rho_s(T)$ were retrieved from the measured Re$V_p(T)$ and the out-of-phase component Im$V_p(T)$ (Fig. 1f). $\rho_s(T)$ is consistent with $\rho_s(T)$ of the best SLCO films synthesized by other research groups[35, 36].

Since the presence of defects and disorder can only lower $\rho_s$, this verifies the high quality of our SLCO films. More details on the synthesis, photolithography and characterizations are included in the Methods section.

To precisely measure both the nematic director and amplitude, we exploited the ARR method[28] and fabricated SLCO films using the pattern shown in Fig. 2a. The current density $\boldsymbol{J}$ at the cross section of the pattern can be controlled to rotate continuously in the *x-y* plane by controlling its *x-* and *y-* components $\boldsymbol{J_x}$ and $\boldsymbol{J_y}$ separately through current injections into contacts $I_{x+}$, $I_{x-}$, $I_{y+}$ and $I_{y-}$. Meanwhile, the corresponding voltages parallel and transverse to $\boldsymbol{J}$, $V_L$ and $V_T$, can be retrieved from the measured $V_x$ and $V_y$ by the simple relations: $V_L = V_x sin\phi + V_y cos\phi$ and $V_T = V_x cos\phi - V_y sin\phi$. Here $\phi$ is the angle between $\boldsymbol{J}$ and the crystalline [100] direction of SLCO film. In this way, we can measure the angular dependences of longitudinal and transverse resistivity $\rho(\phi)$ and $\rho_T(\phi)$ with unprecedented angular resolution ($\Delta\phi < 0.1°$). This is the key for characterization of electronic nematicity because the nematic director in general is not aligned with the lattice, as shown in LSCO[20,22,28]. Thus, the correct characterization of nematicity requires a complete mapping of $\rho(\phi)$ and $\rho_T(\phi)$ for $0 \leq \phi < 360°$. Compared with other schemes, exploiting the vector property and rotating $\boldsymbol{J}$ by controlling its components has the advantage of restricting all the measurements on the same and relatively small area, which greatly reduces the influence of sample inhomogeneity and improves sensitivity. It should be pointed out that the ARR methodology itself does not induce any detectable anisotropy in electric transport. This has been verified by using a gold film as the control sample whose $\rho(\phi)$ and $\rho_T(\phi)$ remain constant for $0 \leq \phi < 360°$ as expected for a normal metal. Detailed discussions on the ARR methodology can be found in our previous publication[28] and references therein.

The longitudinal resistivity $\rho(T)$ taken from the patterned SLCO/KTO film shows the clear superconducting transition (Fig. 2b). The superconducting temperature $T_c$, denoted by the middle point of the transition, is 25.4 K that is close to the values reported in the literature[29], evidencing good crystalline quality and stoichiometry. The angle-dependent $\rho(\phi)$ and $\rho_T(\phi)$ of the normal state from $T = 300$ K down to 30 K manifest clear angular oscillations with 180° period (Figs. 2c and 2d). This illustrates directly that the electric transport is anisotropic along different in-plane directions and the rotational symmetry is spontaneously broken. Note that the resistivity $\rho_T$ transverse to the current direction is strictly zero for normal metals, required by the rotational symmetry of the Fermi liquid. Thus, the nonzero $\rho_T(\phi)$ is a direct measure of the nematic amplitude and comes from the off-diagonal components of the resistivity tensor so that the oscillations in $\rho(\phi)$ and $\rho_T(\phi)$ are correlated through the following expressions:

$$\rho(\phi) = \bar{\rho} + \Delta\rho\cos[2(\phi - \alpha)] \qquad [1]$$

$$\rho_T(\phi) = \Delta\rho\sin[2(\phi - \alpha)] \qquad [2]$$

Here $\bar{\rho}$ is the angle-averaged longitudinal resistivity. $\Delta\rho$ and $\alpha$ are two fitting parameters corresponding to the amplitude and phase offset of the oscillations respectively.

The derivation of the above expressions can be found in our previous publications[20,28]. $\alpha$ represents the angle between the nematic director and the SLCO crystallographic [100] direction. Meanwhile, the unitless nematic amplitude $N$ is determined by $N = \Delta\rho/\bar{\rho}$ that has a geometric meaning similar to the Hall angle.

Using $T$ = 300 K as an example, we show that $\rho(\phi)$ and $\rho_T(\phi)$ oscillations have precise 45° phase difference so that they can be fitted simultaneously with the same $\Delta\rho$ and $\alpha$ using Expressions [1] and [2]. The agreement between the fittings (red curves) and experimental data (black curves) is truly remarkable (see the upper panels in Figs. 2c and 2d). This solidly confirms that the oscillations in $\rho(\phi)$ and $\rho_T(\phi)$ are manifestations of nematicity.

Following the peaks of $\rho(\phi)$ and $\rho_T(\phi)$ oscillations at variable temperatures, the peak angles shift noticeably from 300 K down to 30 K (Figs. 2c and 2d). This rules out the possibility that the anisotropy in electronic transport originates from sample inhomogeneity, structural disorder, or contact misalignment, since all these extrinsic factors are independent of temperature and shouldn't change their orientations with temperature, which is in direct contradiction with the fact that the peak angles shift with temperature. As $T$ further decreases to be close to $T_c$, $\rho(\phi)$ and $\rho_T(\phi)$ oscillations become even stronger in amplitude (Figs. 2e and 2f), implying that the superconducting fluctuations are nematic as well. Moreover, the peak angles shift abruptly as $T$ lowers from 30 K to 26 K, indicating that the nematic director of superconducting fluctuations is different from that of the normal state.

It is of physical significance that the nematic director is not aligned with the SLCO lattice, e.g., $\alpha = 95.84°$ at $T = 300$ K. This implies that the observed nematicity is decoupled from the lattice, and thus its origin is purely electronic. Moreover, the KTO(001) substrate is tetragonal, and due to the nature of epitaxial growth, the SLCO film deposited on top follows the in-plane lattice of KTO. Thus, the SLCO in-plane orthorhombic lattice distortion ratio $\gamma$ ($\gamma = (a - b)/a$, here $a$ and $b$ are the in-plane lattice constants), measured by the X-ray diffraction (XRD) method, is below the instrumental resolution ($\gamma \leqslant 0.05\%$) (Table 1). This, again, from another perspective, verifies that the nematicity in electrical transport is not due to interactions with the lattice but is instead a result of electron-electron correlation. Strictly speaking, due to the electron-lattice coupling, a small lattice distortion should take place, but it is, at best, of secondary importance and cannot be the source of the nematicity.

To elaborate on the effect of lattice distortion, we intentionally chose an orthorhombic substrate $GdScO_3$(110) and deposited SLCO films under the same conditions to facilitate a direct comparison with SLCO/KTO. As confirmed by the XRD measurements, the GSO(110) substrate is orthorhombic with $\gamma = 0.23\%$ and the epitaxially grown SLCO films on top are orthorhombic as well with $\gamma = 0.38\%$ (Table 1). Thus, by comparing the results on these two substrates, we can study the influence of orthorhombic lattice distortion on electronic nematicity. $T_c$ of SLCO/GSO is close to $T_c$ of SLCO/KTO (Fig. 3a), illustrating that the superconductivity is not affected appreciably. The temperature-dependent Hall coefficient $R_H(T)$ taken from SLCO/GSO and SLCO/KTO overlap with each other for $T_c < T \leqslant 300$ K (Fig. 3b), evidencing that the carrier densities in both SLCO films are approximately the same. This is consistent with no change in $T_c$ between SLCO/GSO and SLCO/KTO, since $T_c$ of copper oxide superconductors has a generic dome-shape dependence on the chemical doping level or the carrier density[1,9].

In contrast to $T_c$ and $R_H$, the nematic director of SLCO/GSO differs significantly from that of SLCO/KTO. A unitless quantity, $n(\phi) \equiv \rho_T(\phi)/\bar{\rho}$, is used to represent $\rho_T(\phi)$ normalized by the average longitudinal resistivity $\bar{\rho}$. In polar coordinates, $n(\phi)$ at $T_c$, 30 K, and 300 K show the cloverleaf shape for both SLCO/GSO and SLCO/KTO (Figs. 3c and 3d), intuitively demonstrating the rotational symmetry is spontaneously reduced to the **C₂** symmetry of electronic nematicity. For SLCO/KTO, the orientations of cloverleafs are not aligned with any principal lattice directions and rotate relatively as $T$ cools from 300 K to 30 K and from 30 K to $T_c$. SLCO/GSO, however, is drastically different. There, the nodes of the cloverleafs occur roughly at $\phi = 0°, 90°, 270°, 360°$, which corresponds to the symmetric axes of nematicity—the nematic director. Clearly, the orthorhombic lattice distortion in SLCO/GSO pins the nematic director at all temperatures.

The size of the cloverleafs, which correspond to the nematic amplitude $N$, expands as $T$ cools down to $T_c$. To facilitate a quantitative comparison, we measured both $N(T)$ and $\alpha(T)$ to show the amplitude and director as a function of the temperature (Figs. 3e and 3f). For both SLCO/KTO and SLCO/GSO, $N(T)$ is non-zero even at room temperature and surges when $T$ falls into the narrow temperature range of the superconducting transition. This further strengthens the conclusion that the normal state of SLCO is electronic nematic and the nematicity gets much stronger in the presence of superconducting fluctuations. At 300 K, for both SLCO/KTO and SLCO/GSO, $N(T)$ is about 1.5%, which is close to that of the optimally doped LSCO ($x = 0.16$) with hole doping at the same temperature[20]. For the SLCO normal state, $N(T)$ decreases with increasing temperature (see the inset of Fig. 3e), consistent with the expectation that the nematic order is weakened by thermal fluctuations.

Without orthorhombicity in the lattice, $\alpha(T)$ of SLCO/KTO changes appreciably in the vicinity of $T_c$, indicating that the nematicity has a different orientation in the normal and superconducting states (Fig. 3f). In contrast, $\alpha(T)$ of SLCO/GSO shows that the nematic director is aligned with the SLCO [010] direction for all temperatures. From the XRD measurements on SLCO/GSO, the in-plane lattice constant $a$ is bigger than $b$ (Table 1), meaning the lattice of SLCO is more compressed along the [010] direction. Apparently, the nematic director is pinned by the lattice distortion. Conversely, $N(T)$ of SLCO/KTO is nearly equal to $N(T)$ of SLCO/GSO, implying that the lattice distortion does not enhance or weaken the nematic amplitude. This is one more solid piece of evidence to show that the nematicity is affected but not driven by the lattice distortion.

To examine whether the nematicity is generic to SLCO, we suppress superconductivity by applying an external magnetic field $B$ to expose the hidden ground state of SLCO/KTO at $T < T_c$. As expected, the superconducting state concedes to the normal state as $B$ increases (Fig. 4a). The phase boundaries for the superconducting fluctuating region, which are approximated by the upper and lower bound $T_c^{onset}$ and $T_c^{offset}$, can be determined from the normalized resistivity $\rho(T)/\rho^{50K}$ (Fig. 4c). Concomitantly, the normalized $\rho_T(T)/\rho_T^{50K}$ changes sign and is peaked at $T_p$ as $T$ cools down (Fig. 4b). This is consistent with the rotation of the nematic director $\alpha$ (Figs. 3c and 3f) and the abrupt increase of the nematic amplitude $N$ around $T_c$ (Figs. 3d and 3e). $T_p$ falls between $T_c^{onset}$ and $T_c^{offset}$ for all the $B$ fields and $T_p(B)$ follows the same trend as $T_c(B)$. Thus, as the magnetic field pushes the superconducting fluctuations to lower temperatures, the nematicity peak shifts to lower temperatures accordingly—another unambiguous demonstration of the strong nematicity in superconducting fluctuations.

The complete dependence of the nematicity on temperature and the magnetic field is mapped out in the $T - B$ phase diagram (Figs. 4e and 4f). A clear contrast in $N(T, B)$ and $\alpha(T, B)$ can be observed for the superconducting fluctuating state and the normal state, implying that the origin of the nematicity is different for these two states. Meanwhile, the normal state recovered by the magnetic field at $T < T_c$ shows no appreciable difference in both $N$ and $\alpha$ with the normal state at $T > T_c$. This illustrates that the zero-temperature ground state of SLCO is electronic nematic. Therefore, the unconventional superconductivity of the electron-doped SLCO emerges out of the electronic nematic state.

To examine how electronic nematicity evolves with the carrier density in electron-doped cuprates, we reduced the *in situ* post-growth vacuum annealing time to leave excess interstitial oxygen atoms in the SLCO film so as to effectively reduce the density of mobile carriers. Three SLCO/KTO films were synthesized under the same environment and conditions except for the vacuum annealing time. The sample S3 was annealed at 525 °C for 25 minutes, which is the optimized time to produce optimally doped SLCO. With 15- and 20-minute annealing time for the samples S1 and S2 respectively, the shorter annealing time reduces the effective doping level in SLCO film and drives the SLCO film towards underdoping (Fig. 5a). Compared to S3, the resistivity of S1 rises and the superconducting $T_c$ decreases. Also, the slope of $\rho(T)$ becomes negative at low temperatures for S1. All these are indicative of underdoped cuprates[1, 2]. Despite the differences in annealing time, all three samples remain single crystalline, confirmed by XRD (Fig. 5b). The SLCO(002) peak shifts to lower angles for S1, as compared to S3, evidencing that the *c*-axis lattice constant is larger in S1 (Fig.

5c). This is consistent with the expectation that the interlayer distance between two successive $CuO_2$ planes is expanded by the excess interstitial oxygen atoms in S1.

$n(\phi)$ of S1, S2 and S3 show the cloverleaf shape in the polar coordinates, confirming the presence of electronic nematicity for all these doping levels (Figs. 5d-5f). The rotation of $n(\phi)$ from 300 K to $T_c$ is less significant for the underdoped S1, compared to S3. The rotation of nematic director is also evidenced by the change of the phase offset $\alpha(T)$ close to $T_c$ (Fig. 5g). The increases of $\alpha(T)$ close to $T_c$ are much less in S1 and S2 than in S3. This doping-dependent behavior of nematic director in SLCO is similar to that in the hole-doped LSCO[20]. Moreover, the nematic amplitude $N(T)$ increases as the effective doping level decreases from S3 to S1 (Fig. 5h). The nematicity in both the normal state and superconducting fluctuating state is strengthened with lower doping—another doping-dependent behavior similar to that of the hole-doped cuprate[20].

It should be noted that extrinsic factors, e.g., sample inhomogeneity, thickness variations, or substrate miscut, in principle, can, in principle, induce anisotropy in transport. The anisotropy due to these extrinsic factors occurs randomly from sample to sample and should be independent of the temperature, epitaxial strain, and magnetic field—in direct contradiction to our above experimental facts that the nematicity is sensitively dependent on the temperature, epitaxial strain, and magnetic field. Therefore, these factors can be safely ruled out as possible explanations.

Comparing the electron- and hole-doped[20] cuprates, we see important similarities that unravel universal features intrinsic to high temperature superconductors. Despite the different types of doped carriers, both cuprates, SLCO and LSCO, are electronic nematic in their normal state and superconducting fluctuating state. Rotation of nematic director and enhancement of nematic amplitude take place as long as the normal state transits into the superconducting fluctuating state, even in the presence of an applied magnetic field. The doping dependence of electronic nematicity in SLCO and LSCO is also very similar, i.e., the nematic order is stronger with less doped carriers. A large enough orthorhombic lattice distortion can align the nematic director but not augment the nematic amplitude. These discoveries show the measured nematicity is driven by the electron-electron correlation in both SLCO and LSCO. Thanks to SLCO's lower critical magnetic field than hole-doped cuprates[37], we are able to expose its ground state hidden in the superconducting dome and disclose that it is nematic down to the lowest temperature, while its nematicity matches the nematicity of the normal state above $T_c$ but is significantly different from the nematicity of the superconducting fluctuating state in terms of nematic director and amplitude.

These findings should be contrasted with the fact that the electron- and hole-doped cuprates have critical differences in crystal structure, electronic structure, and phase diagram[1,2]. The electron-doped SLCO has an infinite layer structure with Sr/La atoms intervening between the $CuO_2$ planes. In contrast, the hole-doped LSCO has a perovskite structure consisting of $CuO_6$ octahedra. The doped electrons occupy the Cu site and forms $3d^{10}$ configuration in SLCO, while the doped holes in LSCO occupy the oxygen site and form Zhang-Rice singlets for electrical conduction. Compared to the complicated phase diagram of hole-doped cuprates, filled with abundant electric and spin orders, the phase diagram of electron-doped cuprates is much simpler, e.g., the absence of the pseudo-gap phase[2]. However, despite all these important differences, electronic nematicity behaves in very similar manners for SLCO and LSCO. This implies that the mechanism for electronic nematicity does not depend on the details of structural and electrical properties, but rather is fundamental to cuprates and closely intertwined with high-temperature superconductivity. Our work tightens the restrictions on theoretical modeling of electronic nematicity and invokes in-depth scrutiny on proposed mechanisms[10], e.g., static or dynamic stripe phase[38], and Pomeranchuk instability.

**Methods**

*A. Film synthesis, lithography and characterization*

*1. The $Sr_{0.9}La_{0.1}CuO_2$ thin films on $KTaO_3$(001) substrates*

45 nm (130 unit cells thick) $Sr_{0.9}La_{0.1}CuO_2$ thin films were deposited onto $KTaO_3$(001) substrates by pulsed laser deposition technique. The substrates were degassed at $300°C$ for 20 minutes and then heated to $550°C$ within an oxygen pressure of $5\times10^{-2}$ Torr during film growth. A KrF excimer laser (248 nm) was used with laser intensity 2 J/cm$^2$ and 2 Hz frequency. The growth process was monitored by an *in situ* RHEED system, and the oscillations in the intensities of RHEED spots during growth indicate a nice layer-by-layer growth mode. The post-growth films were cooled to $525°C$ with a rate of 10°$C$/min and held for 25 ~ 30 minutes to remove excess oxygen atoms. This growth recipe results from many rounds of growth-characterization cycles and has been continuously optimized to achieve thin films with the best crystalline quality, the right stoichiometry and high enough $T_c$.

The films were patterned by standard UV lithography and etched by dilute Nitric Acid (DI water: Nitric Acid = 1000:1) for 20s to form the device shown in Fig. 2a, where the Hall bars were 2280 $\mu m$ long and 100 $\mu m$ wide. 50nm gold was deposited onto the contact pads for Ohmic contact.

The in-plane and out-of-plane lattice constants of these films were measured by a high-resolution thin-film X-ray diffractometer (Bruker AXS D8-Discover) with Cu $K\alpha$ radiation ($\lambda = 1.5406$Å).

### 2. The $Sr_{0.9}La_{0.1}CuO_2$ thin films on GdScO$_3$(110) substrates

The synthesis of 28 nm (80 unit cells thick) SLCO on GdScO$_3$(110) substrates is nearly identical to SLCO on KTaO$_3$(001). During growth, the GdScO$_3$(110) substrate temperature was held at 550°$C$ and the laser intensity was 2.1 J/cm$^2$. The post-growth films were cooled to 500°$C$ with a rate of 50°$C$/min and held for 20 ~ 30 minutes. The rest of growth conditions are exactly the same as SLCO on KTaO$_3$(001). The minor changes of growth recipe are necessary for different substrates to achieve the best quality films.

The device fabrication and structural characterization of SLCO films on GdScO$_3$(110) substrates were performed in the same manner as SLCO on KTaO$_3$(001) except that the films were etched by Ar$^+$ ion milling.

### B. Transport measurements

The samples were mounted in a cryocooler to reach temperatures as low as 4 K and magnetic fields up to 1 T for Hall effect measurements. For magneto-transport measurements under higher magnetic fields, the samples were loaded into a Physical Property Measurement System (Quantum Design, PPMS 16T) to reach 1.8 K and 16 T magnetic field. For the ARR measurements, two Keithley 2450 soucemeters were used to independently control $I_x$ and $I_y$; meanwhile, two Keithley 2182A nanovoltmeters were used to measure $V_x$ and $V_y$ simultaneously.

## C. Two-coil mutual inductance measurements

The sample is inserted into a gap between the drive coil and the pick-up coil. The drive coil has 540 turns of gauge 15 μm copper wires, and the pick-up coil has 510 turns of 15 μm copper wires. The inner/outer diameters of the drive and pick-up coils are 0.42/1.85 mm and 0.65/2.1 mm, respectively. The distance between the centers of the two coils is 1.22 mm. During measurement, a 98.94 μA *ac* current with 30 kHz frequency is generated by a Stanford Instruments SR830 DSP Lock-in amplifier and is connected to the drive coil, generating an *ac* magnetic field around 1 Gauss, which is well below the critical field of SLCO films. The induced voltage in the pick-up coil is fed to the Lock-in amplifier to measure both its real and imaginary components. The complete setup is mounted onto a cryocooler to reach temperatures as low as 4 K.

**Data Availability**

The data that support the findings of this study and all other relevant data are available from the corresponding author upon request.

**Acknowledgements**

This work was supported by Research Center for Industries of the Future (RCIF project No. WU2023C001 to J.W.) at Westlake University, the National Natural Science Foundation of China (Grant No. 12174318 to J.W.), the Zhejiang Provincial Natural

Science Foundation of China (Grant No. XHD23A2002 to J.W.). We acknowledge the assistance provided by the Westlake Center for Micro/Nano Fabrication and the Instrumentation and Service Center for Physical Sciences at Westlake University.

**Author contributions**

The films were synthesized and characterized by J.Y.S., the lithography was done by J.Y.S. and Y.C.Z., the transport measurements were done by J.Y.S., G.F.C., and G.Y.X., the mutual inductance measurements were done by J.Y.H. and X.B.C., and the analysis and interpretation put forward by J.W.

**Competing interests**

The authors declare no competing interests.

**Additional information**

Correspondence and requests for materials should be addressed to J.W.

|  | *a* | *b* | *c* | *γ* |
|---|---|---|---|---|
| **KTO(001)** | 3.989Å | 3.989Å | 3.989Å | 0 |
| **SLCO(001)/KTO(001)** | 3.981Å | 3.979Å | 3.402Å | 0.05% |
| **GSO(110)** | 3.974Å | 3.965Å | 3.974Å | 0.23% |
| **SLCO(001)/GSO(110)** | 3.957Å | 3.942Å | 3.403Å | 0.38% |

**Table 1 │ The orthorhombic lattice distortion ratio *γ* of KTO, GSO substrates and SLCO films grown on top, determined by XRD measurements.** $\gamma = (a-b)/a$, here *a* and *b* are in-plane lattice constants. *c* is the out-of-plane lattice constant.

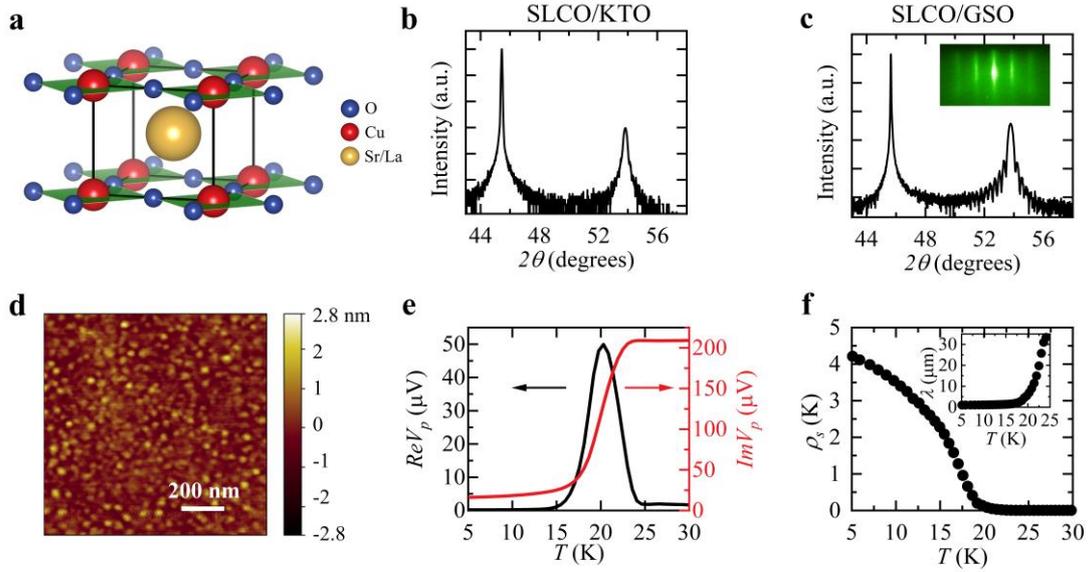

**Figure 1 | Structural characterization of synthesized SLCO films. a,** A schematic drawing of SLCO lattice structure. **b**, and **c**, The *c*-axis X-ray diffraction (XRD) spectrum for the SLCO (45 nm)/KTO and SLCO (28 nm)/GSO films respectively. The peak ($2\theta \sim 53.8°$) corresponds to the SLCO film, demonstrating it is single crystal. The inset in panel **c** is the reflection high energy electron diffraction (RHEED) pattern taken on the post-annealing SLCO/GSO film, showing good crystalline quality and morphology. **d**, An AFM image taken on SLCO (35 nm)/KTO. The r.m.s. surface roughness is 0.44 nm. **e**, The diamagnetic signal (Meissner effect) of SLCO measured by the two coil mutual inductance method. $\mathrm{Re}V_p$ and $\mathrm{Im}V_p$ are the real (in phase) and imaginary (out of phase) voltages of the pick-up coil. **f**, The temperature dependence of the superfluid stiffness $\rho_s$ and the penetration depth $\lambda$ (the inset).

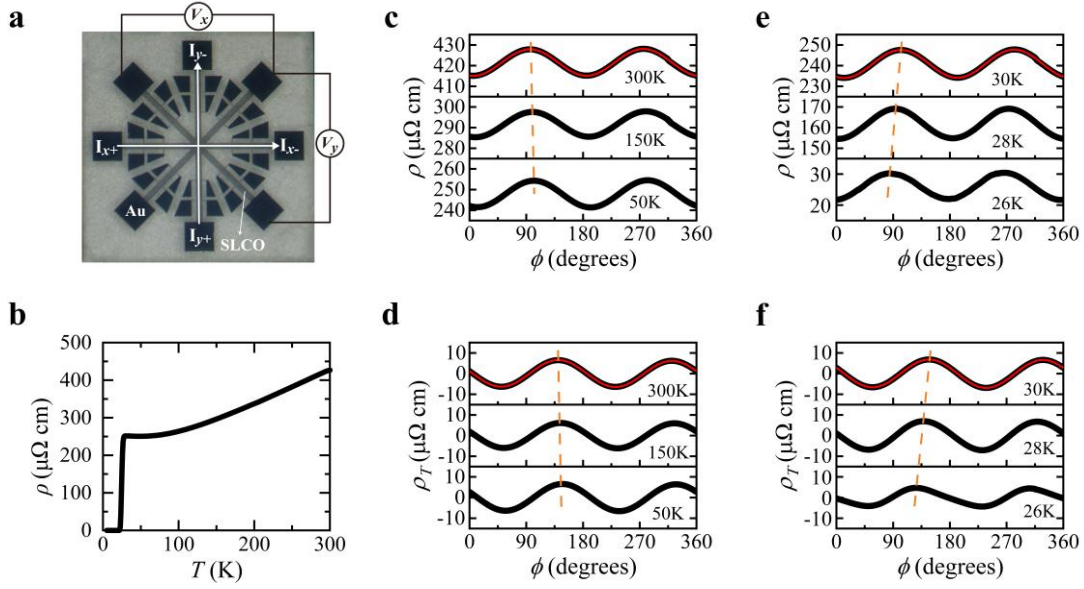

**Figure 2 | Electronic nematicity in SLCO/KTO measured by the ARR method. a,** A photo of SLCO/KTO sample patterned for conduction of the ARR measurements. The grey area is the patterned SLCO film, and the dark square pads are covered with gold for electric contacts. The current direction at the center of the cross section is rotated in-plane continuously by controlling its $x$- and $y$- components $I_x = I_0 cos\phi$ and $I_y = I_0 sin\phi$, where $\phi$ is the angle between the current and the SLCO [100] direction. The longitudinal and transverse resistivity, $\rho(\phi)$ and $\rho_T(\phi)$, can be retrieved from the measured $V_x(\phi)$ and $V_y(\phi)$. **b,** $\rho(T)$ shows clear superconductivity. $T_c$ is defined as the middle point of the superconducting transition and $T_c \sim 25.4$ K, in agreement with literature reports[29]. **c,** and **d,** $\rho(\phi)$ and $\rho_T(\phi)$ at three representative temperatures, 300 K, 150 K, and 30 K, manifest appreciable angular oscillations with 180° period, evidencing the SLCO normal state is nematic. The angles corresponding to the peaks in $\rho(\phi)$ and $\rho_T(\phi)$, indicated by the dashed line, are shifting with temperature, showing the change of the nematic director in the normal state. **e,** and **f,** The same as **c** and **d**, except for temperatures in the vicinity of $T_c$. The shift in the peak angles illustrates the nematic director is altered by superconducting fluctuations. The red curves in **c-f** are fittings according to Expressions [1] and [2]. The agreements between the fittings and the data (black curves) are truly remarkable.

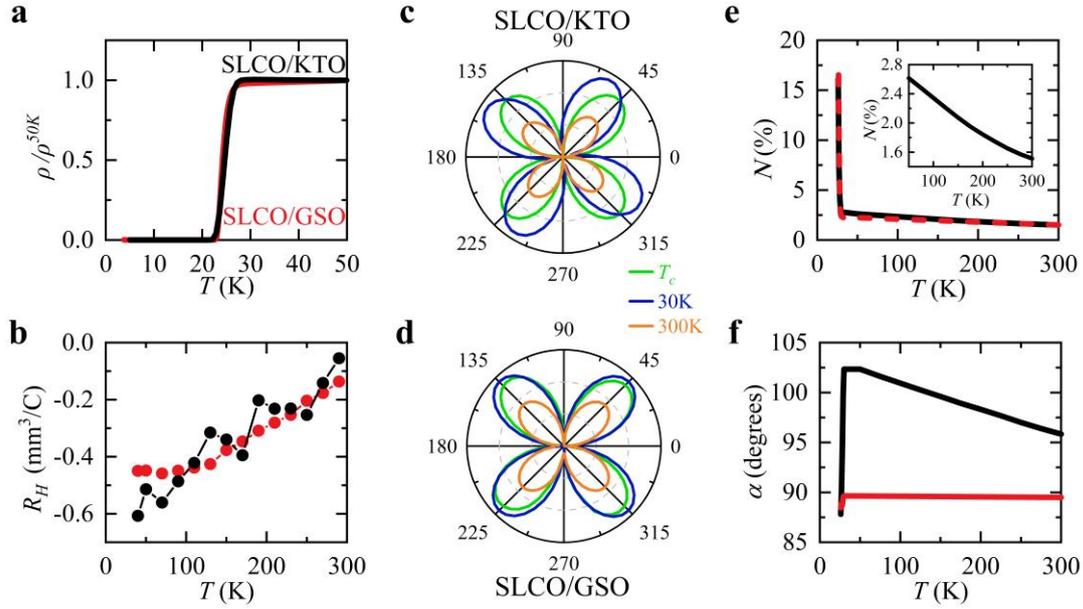

**Figure 3 │ The effect of orthorhombic lattice distortion on nematicity in a tetragonal SLCO/KTO film and an orthorhombic SLCO/GSO film. a,** and **b,** $T_c$ and the Hall coefficient $R_H(T)$ of SLCO/KTO and SLCO/GSO are roughly the same, showing superconductivity and the carrier density are not affected by the lattice distortion. The thickness of SLCO/KTO is 45 nm and SLCO/GSO is 28 nm. **c,** $n(\phi) \equiv \rho_T(\phi)/\bar{\rho}$, denotes the normalized angular oscillations. For SLCO/KTO, $3 \times n(\phi)$ at 300 K (orange), $3 \times n(\phi)$ at 30 K (blue) and $n(\phi)$ at $T_c$ (green) manifest clear **C₂** symmetry under the polar coordinate system. Here, $n(\phi)$ at 300 K and 30 K are amplified by three times for clear visibility. **d,** The same, except for SLCO/GSO. The orientation of $n(\phi)$ is aligned with the lattice at all three temperatures, in stark contrast to that in **c**. This shows the nematic director is pinned by the lattice distortion in SLCO/GSO. **e,** and **f,** The nematic amplitude and director, $N(T)$ and $\alpha(T)$, show drastic differences for $T > T_c$ and $T \sim T_c$. Results for SLCO/KTO and SLCO/GSO are colored in black and red respectively. $N(T)$ for SLCO/GSO is denoted by the dashed line for better visibility. The inset of panel **e** is the magnified view of $N(T)$ of the SLCO/KTO normal state, which is strongly temperature dependent.

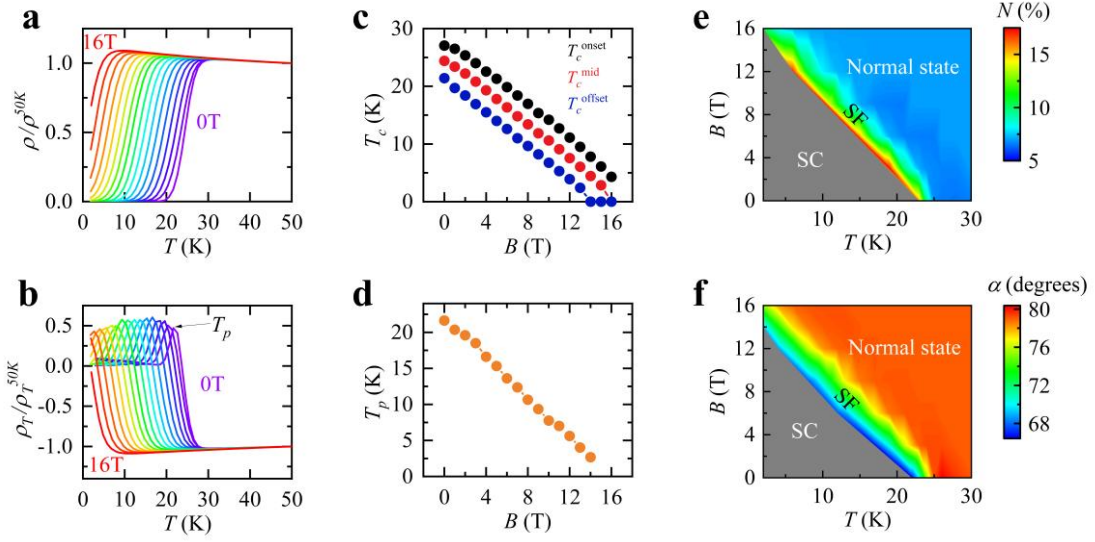

**Figure 4 │ The *B* - *T* phase diagram of electronic nematicity in SLCO/KTO. a,** The superconductivity is suppressed step by step by the applied magnetic field *B*, shown by the normalized resistivity $\rho(T)/\rho^{50K}$ (corresponds to $B$ = 0 to 16 T in 1 T step from bottom to top). **b,** The normalized $\rho_T(T)/\rho_T^{50K}$ manifests a pronounced peak at $T_p$ and $T_p$ shifts to lower temperature in accordance with $\rho(T)/\rho^{50K}$ as *B* increases, implying that the peak in $\rho_T/\rho_T^{50K}$ traces the superconducting transition. For **a** and **b**, the current is applied along SLCO [100] direction. **c,** $T_c^{onset}$, $T_c^{mid}$, and $T_c^{offset}$, which corresponds to 10%, 50% and 90% drop respectively in the normal state $\rho(T)$, are used to denote the temperature range with strong superconducting fluctuation. **d,** $T_p(B)$ falls into this temperature range and traces $T_c(B)$ closely, unambiguously showing its direct correlation with superconductivity. **e,** and **f,** *B* - *T* phase diagram of the nematic amplitude *N* and director *α*. "SC" stands for the superconducting state and "SF" for the superconducting fluctuating state.

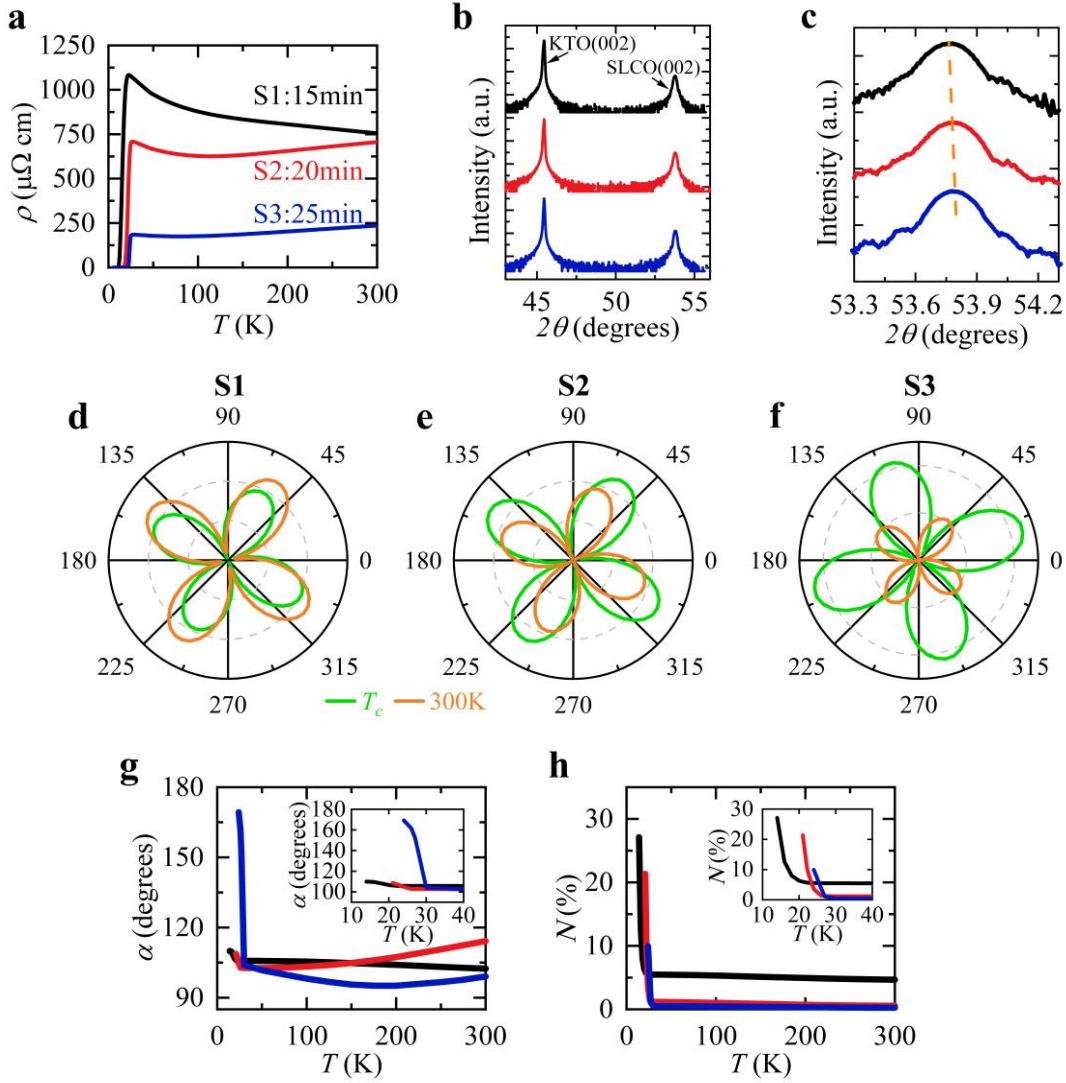

**Figure 5 | The electronic nematicity of underdoped SLCO films. a**, The effective doping level in SLCO can be controlled by *in situ* post-growth vacuum annealing. For SLCO samples S1, S2 and S3, they are synthesized under identical conditions except that the annealing time are 15, 20 and 25 minutes respectively. The shorter annealing time for S2 and S3 leaves excess interstitial oxygen atoms in SLCO, which reduces the iterant electron density. **b**, The XRD spectrums for S1, S2 and S3, confirming all three films are single crystalline. S1, S2 and S3 are consistently colored in black, red, and blue respectively for all panels in this figure. **c**, The magnified view of SLCO (002) XRD peaks. The diffraction peak shifts to higher angles from S1 to S3, indicated by the dashed line, implying that the *c*-axis lattice constant increases as the doping level of SLCO films lowers towards underdoping. **d**, **e**, and **f**, $n(\phi)$ at $T_c$ (green) and $2 \times n(\phi)$ at 300 K (orange) for S1, S2 and S3 respectively. **g**, and **h**, The nematic amplitude and

phase offset, $N$ and $\alpha$, for S1, S2 and S3. The insets are the magnified views of the low temperature part.